\documentclass[twocolumn]{aastex61}
\pdfoutput=1
\usepackage[latin1]{inputenc}
\usepackage{amsmath}
\usepackage{amsfonts}
\usepackage{amssymb}
\usepackage{chngcntr}

\usepackage{graphicx}
\usepackage[varg]{txfonts}
\usepackage{mwe}  
\usepackage{url}
\usepackage{longtable}
\usepackage{float}
\shorttitle{CXB and AGN Parameter Space}
\shortauthors{Ananna et al.}

\begin{document}
\title{Accretion History of AGN II: Constraints on AGN Spectral Parameters using the Cosmic X-ray Background}
	
\author{Tonima Tasnim Ananna}
\affiliation{Department of Physics \& Astronomy, Dartmouth College, 6127 Wilder Laboratory, Hanover, NH 03755, USA}
\affiliation{Department of Physics, Yale University, PO BOX 201820, New Haven, CT 06520-8120}
\affiliation{Yale Center for Astronomy and Astrophysics, P.O. Box 208120, New Haven, CT 06520, USA 0000-0002-5554-8896}
\email{tonimatasnim.ananna@yale.edu}

\author{Ezequiel Treister}
\affiliation{Instituto de Astrof\'{\i}sica, Facultad de F\'{\i}sica, Pontificia Universidad Cat\'{o}lica de Chile, Casilla 306, Santiago 22, Chile}

\author{C. Megan Urry}
\affiliation{Department of Physics, Yale University, PO BOX 201820, New Haven, CT 06520-8120}
\affiliation{Yale Center for Astronomy and Astrophysics, P.O. Box 208120, New Haven, CT 06520, USA 0000-0002-5554-8896}
\affiliation{Department of Astronomy, Yale University, P.O. Box 208101, New Haven, CT 06520, USA}

\author{C. Ricci}
\affiliation{Instituto de Astrof\'{\i}sica, Facultad de F\'{\i}sica, Pontificia Universidad Cat\'{o}lica de Chile, Casilla 306, Santiago 22, Chile}
\affiliation{Kavli Institute for Astronomy and Astrophysics, Peking University, Beijing 100871, China}

\author{Ryan C. Hickox}
\affiliation{Department of Physics \& Astronomy, Dartmouth College, 6127 Wilder Laboratory, Hanover, NH 03755, USA}

\author{Nikhil Padmanabhan}
\affiliation{Department of Physics, Yale University, PO BOX 201820, New Haven, CT 06520-8120}
\affiliation{Yale Center for Astronomy and Astrophysics, P.O. Box 208120, New Haven, CT 06520, USA 0000-0002-5554-8896}
\affiliation{Department of Astronomy, Yale University, P.O. Box 208101, New Haven, CT 06520, USA}

\author{Stefano Marchesi}
\affiliation{INAF - Osservatorio di Astrofisica e Scienza dello Spazio di Bologna, Via Piero Gobetti, 93/3, 40129, Bologna, Italy}
\affiliation{Department of Physics and Astronomy, Clemson University,  Kinard Lab of Physics, Clemson, SC 29634, USA}

\author{Allison Kirkpatrick}
\affiliation{Department of Physics \& Astronomy, University of Kansas, Lawrence, KS 66045, USA}

\begin{abstract}
   We constrain X-ray spectral shapes for the ensemble of AGN based on the shape of the Cosmic X-ray Background (CXB).
   Specifically, we rule out regions of X-ray spectral parameter space that do not reproduce the CXB in the energy range 1$-$100 keV. The key X-ray spectral parameters are 
   the photon index, $\Gamma$; the cutoff energy, $E_{\rm cutoff}$; and the reflection scaling factor, $R$. Assuming each parameter follows a Gaussian distribution, we first explore the parameter space using a Bayesian approach and a fixed X-ray luminosity function (XLF). For $\sigma_E=36$ keV and $\sigma_R=0.14$, fixed at the observed values from the \textit{Swift}-BAT 70-month sample, 
   we allow $\langle R\rangle$, $\langle E_{\rm cutoff}\rangle$ and $\langle \Gamma\rangle$ to vary subject to reproducing the CXB. We report results for $\sigma_\Gamma=0.1-0.5$. In an alternative approach, we define the parameter distributions, then forward model to fit the CXB by perturbing the XLF using a neural network. This approach allows us to
   rule out parameter combinations that cannot reproduce the CXB for any XLF. The marginalized conditional probabilities for the four free parameters are: $\langle R \rangle=0.99^{+0.11}_{-0.26}$, $\langle E_{\rm cutoff}\rangle=118^{+24}_{-23}$, $\sigma_\Gamma=0.101^{+0.097}_{-0.001}$ and $\langle \Gamma\rangle=1.9^{+0.08}_{-0.09}$. We provide an interactive online tool 
   for users to explore any combination of $\langle E_{\rm cutoff} \rangle$, $\sigma_E$, $\langle \Gamma \rangle$, $\sigma_\Gamma$, $\langle R \rangle$ and $\sigma_R$, including different distributions for each absorption bin, subject to the integral CXB constraint.
   The distributions observed in many AGN samples can be ruled out by our analysis, meaning these samples can not be representative of the full AGN population.
   The few samples that fall within the acceptable parameter space are hard X-ray-selected, commensurate with their having fewer selection biases.
\end{abstract}


\section{Introduction}\label{sec:intro}

An accreting supermassive black hole (SMBH) appears brightly in most wavelengths of the electromagnetic spectrum. X-rays are considered to be one of the most unbiased tracers of active galactic nuclei (AGN) as they are produced very close to the SMBH - in the corona of the accretion disk - and they can penetrate heavily obscuring column densities. 
\citeauthor{ananna2018} (\citeyear{ananna2018}; henceforth Paper~I) presented a comprehensive model of SMBH growth in the form of an evolving X-ray luminosity function (XLF). XLFs describe the underlying AGN population as number density of AGNs at each epoch/redshift z, as a function of intrinsic luminosity L and obscuration N$_{\rm H}$, quantified in terms of equivalent hydrogen column density of obscuring material. A population synthesis model combines an XLF and a set of AGN X-ray spectra to reproduce observed X-ray constraints (e.g.,  \citealp{maccacaro1991,madau1994,comastri1995,ueda2003,gilli2007,akylas2012,shi2013,ueda2014,aird2015xlf}, Paper~I).

A correct population synthesis model is able to reproduce constraints such as the Cosmic X-ray background (CXB), which describes the total observed intensity from X-ray sources at a given energy level per unit solid angle of the sky. The model prediction of the CXB at each energy value is calculated by integrating over the product of AGN spectra and space densities, as follows:
\begin{multline}
CXB_{\rm model} (E) \propto \int^{z_{\rm max}}_{z_{\rm min}} \int^{L_{\rm max}}_{L_{\rm min}} \int^{\log N_{\rm H, max}}_{\log N_{\rm H, min}}
XLF(z, L, \log N_{\rm H} )~\times \\
{\rm Spectra}(E*[1+z])~\times \frac{dV}{dz}~dz~dL~d \log N_{\rm H} .
\end{multline}
The {\citet{gilli2007} population synthesis model explored some parts of AGN spectral parameter space and showed that, independent of the underlying XLF, certain AGN spectral parameter combinations cannot reproduce the CXB consistently at all energies. Similarly, \citet{akylas2012} systematically explored the degeneracy between the XLF and two parameters of the AGN spectra.
However, a full systematic exploration of AGN parameter space 
has not been undertaken using population synthesis models. 
The 
population synthesis model in Paper~I fit several dozen X-ray constraints, and a neural network simultaneously ensured a good fit to the CXB.
Here, we present a more comprehensive exploration of the AGN spectral parameter space using this population synthesis model, in conjunction with recently available X-ray spectral models that are more realistic and sophisticated (\S~2), as well as a neural network that can perturb the XLF in an automated way.

Specifically, in this work (see also \S~6 of Paper~I), we show that the quality of fit to the CXB is determined by the relative contributions in different absorption bins. AGN obscuration has three classes: unobscured objects are AGN shrouded in an equivalent hydrogen column density of  N$_{\rm H}$ $<$ 10$^{22}$ cm$^{\rm -2}$, Compton-thin objects have 10$^{22}$ cm$^{\rm -2}$ $\leq$ N$_{\rm H}$ $<$ 10$^{24}$ cm$^{\rm -2}$ and Compton-thick objects have N$_{\rm H}$ $\geq$ 10$^{24}$ cm$^{\rm -2}$. The summed contribution of AGNs over all N$_{\rm H}$ produces the observed CXB. We show in \S~\ref{sec:spectra} (see also Figure~8 of Paper~I) that for some spectral parameters, including values observed in some AGN surveys, the CXB cannot be reproduced regardless of the underlying XLF. We discuss the selection biases and spectral modeling difficulties in survey data which might lead to this issue in \S~\ref{sec:spectra}.

Additionally, there are some gaps in the literature about the limits of spectral parameter spaces that can reproduce the CXB. Usually, AGN spectral parameters in population synthesis models are determined through observed distributions in surveys (e.g., \citealp{ueda2014,buchner2015}). In Figures~11 and 18 of Paper~I, we showed that several previous population synthesis models chose spectral parameter sets that do not reproduce the CXB well when coupled with the XLFs presented in those works. Therefore an XLF-independent exploration of AGN parameter space will provide insights about its acceptable regions when constructing population synthesis models, in addition to providing a reference point for comparison with observations from samples selected using different methods.

In this work, we use the most updated X-ray spectral models to construct AGN spectra. We explore the distribution of spectra that can reproduce the CXB using two approaches: the first approach assumes that our current observationally constrained XLF is correct, and the parameter space of AGN spectra can be constrained by exploring the quality of the fit to the CXB when coupled with this XLF. 
For the second approach, we assume a spectrum rather than an XLF, and forward model using the neural network to find { the XLF that best fits} the CXB given the assumed spectrum. 
We apply the second approach systematically across parameter space, and use the quality of the fit to reject regions that fail to reproduce the CXB for any XLF. We also show how the allowed region shifts as key spectral parameters change. Finally, we provide an online interactive tool with which users can explore the goodness of fit to the CXB for any combination of spectral parameters.

\begin{figure*}[th]
	\centering
	\includegraphics[width=1.0\linewidth]{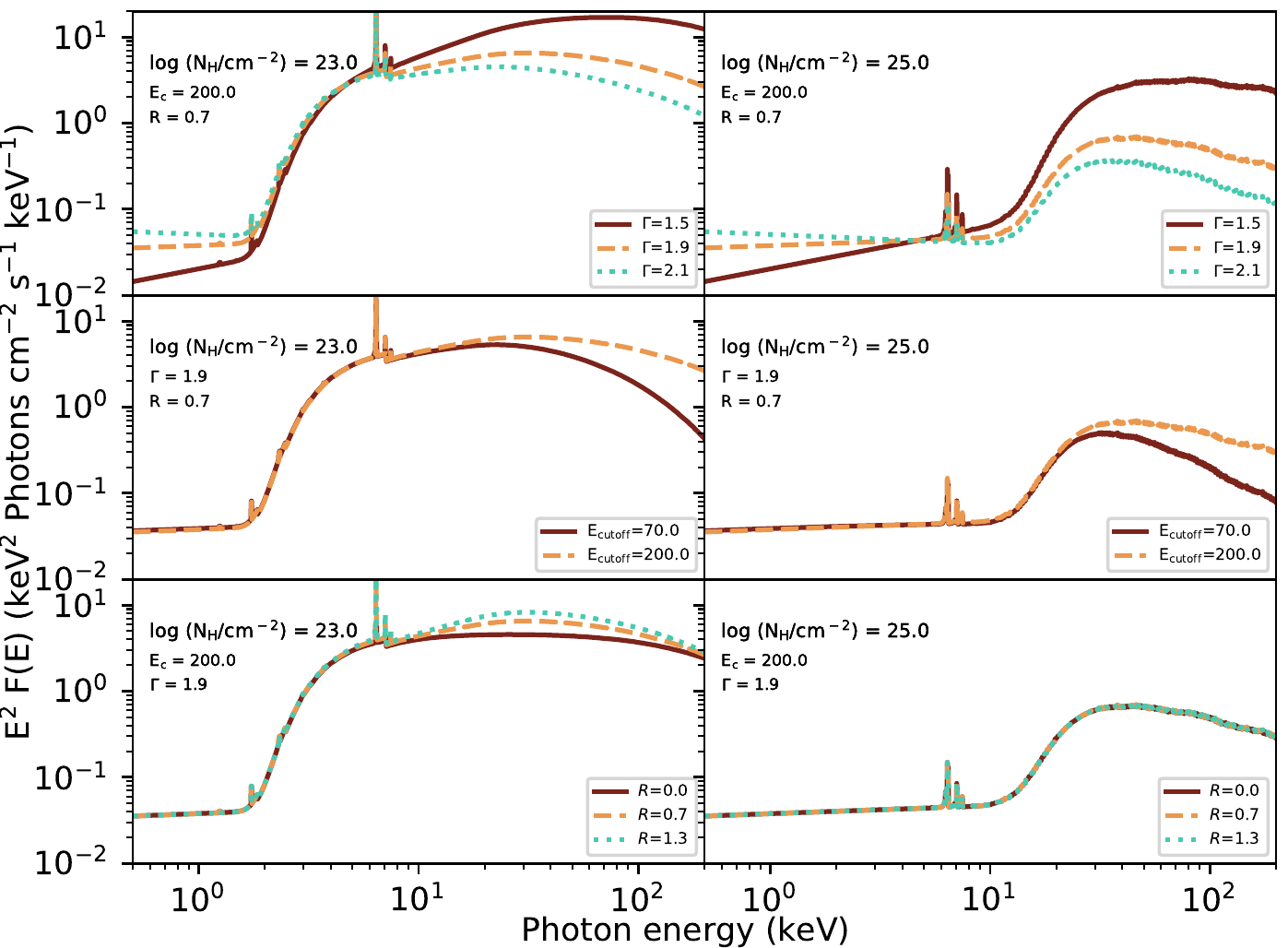}
	\caption{Dependence of X-ray spectral shape on key parameters, for two average column densities of the obscuring torus: $\log$ (N$_{\rm H}/{\rm cm}^{\rm -2}$) = 23 (\textit{left panels}) and  $\log$ (N$_{\rm H}/{\rm cm}^{\rm -2}$) = 25 (\textit{right panels}). \textit{Top panels} show three values of the power-law index, $\Gamma$=1.5 (\textit{solid red lines}), 1.9 (\textit{dashed orange lines}), and 2.1 (\textit{dotted green lines}); the other spectral parameters are fixed, with scattering fraction, f$_{\rm scatt}$=1\%, cutoff energy, E$_{\rm cutoff}$ = 200 keV, and reflection scaling factor, R = 0.7. 
	\textit{Middle panels} show two values of the cutoff energy, E$_{\rm cutoff}$=70~keV (\textit{solid red lines}) and 200~keV (\textit{dashed orange lines}), with f$_{\rm scatt}$=1\%, $\Gamma$ = 1.9, and R = 0.7. 
	\textit{Bottom panels}: three values of the reflection fraction, R=0.3 (\textit{solid red lines}), 0.7 (\textit{dashed orange lines}) and 1.0 (\textit{dotted green lines}), with fixed f$_{\rm scatt}$=1\%, $\Gamma$ = 1.9, and R = 0.7. Details about the \textsc{xspec} models are given in \S~\ref{ssec:spectra_method}. 
	} 
	\label{fig:gamma_variation_spectra} 
\end{figure*} 

In this paper, we present our work as follows:  \S~\ref{sec:spectra} describes the key AGN spectral parameters and their values as derived from survey samples. In \S~\ref{sec:method} we describe how we vary the components of the population synthesis model while still fitting the Cosmic X-ray Background. \S~\ref{sec:results} describes our results, namely, the unacceptable spectral shapes, and \S~\ref{sec:conclusion} presents discussion and conclusions.

\begin{deluxetable*}{lllrll}[th]
\tablewidth{0pt}
\tablecaption{\label{tab:spectral_parameters_in_papers}\textsc{Observed X-Ray Spectral Parameters of AGN}}
 \tablehead{& \colhead{\textsc{Sample}} & \colhead{\textsc{Photon Index}} & \colhead{\textsc{Cutoff Energy}} & \colhead{\textsc{Refl. Scaling Factor}} & \\
& &  \colhead{\textsc{$\Gamma$\tablenotemark{1}}} & \colhead{\textsc{E$_{\rm cutoff}$ (keV)\tablenotemark{1}}} & \colhead{\textsc{R}\tablenotemark{1}} }
\startdata
\citet{claudio2017bat} & \textit{Swift}-BAT 14$-$195 keV selected (local) & 1.80$^{+0.02}_{-0.02}\pm$0.17, 1.76$^{+0.02}_{-0.02}\pm$0.17 \tablenotemark{2} & 210$\pm$36, 188$\pm$27 & 0.83$\pm$0.14, 0.37$\pm$0.11 \\
\citet{zap2018} & \textit{NuSTAR} 8$-$24 keV selected & 1.89$\pm$0.26\tablenotemark{3} & 200 (fixed) & 0.67, 0.28\tablenotemark{4} \\
\citet{malizia2014} & INTEGRAL 0.3$-$100 selected type-1 & 1.73$\pm$0.17 & 128$\pm$46\tablenotemark{5} &  \\
\citet{buchner2014spectra} & 0.5$-$8 keV selected (CDFS 4Ms) & 1.9$-$2.0 & NA & 0.7$\pm$0.5 \tablenotemark{6}  \\
\citet{bntorus2011} & IR-selected type-1 and type-2 & 1.90$^{+0.05}_{-0.07}\pm$0.31$^{+0.05}_{-0.05}$ & NA & NA \\
\citet{scott2011} & Type 1 (optically selected \textit{XMM} spectra) & 1.99$^{+0.01}_{-0.01} \pm 0.3^{+0.01}_{-0.01}$ & &  \\
\citet{beckmann2009} & Hard X-ray selected ($>$ 20 keV) & 1.96$\pm$0.02, 1.91$^{+0.02}_{-0.03}$ & 86$^{+21}_{-14}$, 184$^{+16}_{-52}$ & 1.2$^{+0.6}_{-0.3}$,1.1$^{+0.7}_{-0.4}$\tablenotemark{7} \\
\citet{dadina2008} & 2-100 keV selected (local) & 1.89$\pm$0.03, 1.80$\pm$0.05 & 230$\pm$22, 376$\pm$42 & 1.23$\pm$0.11, 0.87$\pm$0.14 \\
\citet{derosa2012} & INTEGRAL selected ($>$ 20 keV) & 1.80$\pm$0.18,1.75$\pm$0.26 & 73$\pm$25 & 1.63$\pm$1.15 \tablenotemark{8} \\
\citet{ueda2014} & \textit{Swift}-BAT 9-month sample\tablenotemark{9} & 1.94$^{+0.03}_{-0.03}\pm$0.09$^{+0.05}_{-0.05}$, 1.84$^{+0.04}_{-0.04}\pm$0.15$^{+0.06}_{-0.06}$ & 300 & 0.5 \\
\enddata
\tablenotetext{1}{For unobscured and obscured sources, respectively. The subscripts and superscripts on the values are standard errors on the means and dispersions of the distributions.}
\tablenotetext{2}{Compton-thick objects have $\Gamma = 2.05 \pm 0.17$, E$_{\rm cutoff} = 449 \pm 64$ and R = 0.15$\pm$0.12. The inclination angle of all objects was fixed at 30$^{\circ}$.}
\tablenotetext{3}{The distribution of the measured $\Gamma$ peaks between 1.8 and 2.0, with a mean at 1.89.}
\tablenotetext{4}{The interquartile ranges(25-75\%) for unobscured and obscured sources are 0.10$-$1.8 and 0.05$-$1.07, respectively.}
\tablenotetext{5}{Note that this is mean cutoff energy for which this quantity could be constrained, i.e., ignoring lower limits. The baseline model used for fitting uses \textsc{pexrav}; however, no reflection scaling factor was reported.}
\tablenotetext{6}{From Figure~10 of \citet{buchner2014spectra}.}
\tablenotetext{7}{For these spectra, the cutoff energy can only be determined when not fitting the spectra with a reflection component. Also, the inclination angles assumed for Type-1 and Type-2 AGN are 30$^{\circ}$ and 60$^{\circ}$, respectively.}
\tablenotetext{8}{Correlated with $\Gamma$, so that a higher $\Gamma$ produces a higher value of R. The cutoff energy listed is for the sample of 10 objects for which this quantity could be constrained.}
\tablenotetext{9}{The spectra for the \citet{ueda2014} XLF were constrained using \textit{Swift}-BAT 9-month catalog. However, the space densities were constrained using several other surveys, as detailed in that work.}
\end{deluxetable*}

\section{Observed AGN spectral parameters}\label{sec:spectra}

\noindent
AGN have an accretion disk surrounding the SMBH, which is the source of emission in optical and UV bands (\citealp{ss1973,shields1978,risaliti2004}). Above the accretion disk is a hot corona which emits at X-ray wavelengths. This intrinsic X-ray spectrum is a power-law, with a photon index in the range $\Gamma \simeq$ 1.4 - 2.1 (\citealp{nandra1994,ueda2014,claudio2017bat} - more references in Table~\ref{tab:spectral_parameters_in_papers}) and an exponential cutoff at high energies, i.e., F(E) $\propto E^{-\Gamma} \exp(-E/E_{\rm cutoff})$. 
This power-law emission is reflected by the accretion disk; in \textsc{xspec} \citep{xspec} the reflected component is often modeled by \textsc{pexrav} or \textsc{pexmon} \citep{pexrav,pexmon}, with an adjustable scaling factor, R. 
According to AGN unification, the SMBH and accretion disk likely reside inside a torus of gas and dust 10$-$100 pc away from the SMBH. High energy X-rays can penetrate the dusty torus, revealing the AGN within. The transmitted X-ray spectrum can be modeled using \textsc{mytorus} \citep{mytorus} or \textsc{borus02} \citep{mislav2018}. The unabsorbed continuum is scattered by gas outside the torus region, within the cone of its opening angle. The dependence of the spectral shape on $\Gamma$, E$_{\rm cutoff}$ and R is shown in Figure~\ref{fig:gamma_variation_spectra}, for two different absorbing column densities, log (N$_{\rm H}$/cm$^{-2}$) = 23 and 25.

Often, AGN parameter space is constrained by fitting X-ray spectra of AGN detected in surveys. However, AGN surveys are subject to selection biases, so that the observed distribution of parameters may not reflect the underlying distribution in nature. Additionally, accurate spectral fits are only possible for sources that have a lot of counts.
Table~\ref{tab:spectral_parameters_in_papers} lists observed AGN parameters for different samples.

\citet{claudio2017bat} published a detailed analysis of both obscured and unobscured AGN in the local Universe from the \textit{Swift}-BAT 70-month survey, one of the least biased AGN surveys. They report that for \textit{Swift}-BAT observed spectra, constraining certain parameters, such as reflection scaling factor R, is easier when the value is high (R $>$ 1), whereas cutoff energy is well constrained only when the value is low (E$_{\rm cutoff} < 100$ keV). For the rest of the objects, one can only obtain limits on these parameters. \textit{Swift}-BAT is sensitive in the 14$-$195 keV energy range; for lower energy bands, such as the $0.1-8$ keV ($0.1-10$ keV) band of \textit{Chandra} (\textit{XMM}), constraining the reflection scaling factor or E$_{\rm cutoff}$ is more difficult, especially at lower redshifts. This is another reason the observed distributions of parameters do not necessarily reflect the intrinsic ones. Additionally, spectral parameters can be coupled. \citet{z99} and \citet{petrucci2001} reported a correlation between reflection parameter and photon index, \citet{matt2001} reported a positive correlation between photon index and cutoff energies. These correlations may occur due to the intrinsic nature of the spectra, or due to the fact that these parameters are strongly related in the fitting procedure \citep{akylas2012}. Similarly, the difference of observed parameters between obscured and unobscured sources might be intrinsic but can also arise due to imperfections in the modeling of the obscurer (\citealp{elitzur2012,mislav2018}). 

When modeling AGN spectra in population synthesis models, usually a single spectrum, i.e., without any dispersion in values of parameters, or a set of spectra (i.e., incorporating some distribution/dispersion of parameters) is assumed. An XLF provides the space densities of AGN, which integrated together with the spectra, can reproduce the CXB and number counts.
The population synthesis model of \citet{gilli2007} was the first to use a distribution of photon indices for calculating the CXB, rather than a single value; this tended to harden the CXB peak emission at $E\sim10$~keV. Subsequent XLFs also assumed some dispersion in the values of some of these parameters; \citet{ueda2014} assumed a single R and E$_{\rm cutoff}$, but a dispersion in $\Gamma$, and \citet{aird2015xlf} assigned distributions for both $\Gamma$ and R and kept E$_{\rm cutoff}$ constant. In Paper~I, we assigned dispersions to all three parameters. However, we note that any given XLF can only reproduce the CXB for a certain range in spectral parameter space.

\citet{gilli2007} presented an XLF where the unobscured and Compton-thin space densities were constrained using surveys, and the space densities of Compton-thick objects were constrained using the residual X-ray background. 
They reported that with $\langle \Gamma \rangle = 1.9, \sigma_{\Gamma} = 0.2$ and a $\langle $E$_{\rm cutoff} \rangle$ = 300 keV, either the peak CXB (at E = 30 keV) is underestimated and the CXB at E = 100 keV is well-fitted, or the CXB at E = 30 keV is well-fitted but at E = 100 keV is overestimated, depending on how many Compton-thick sources are added. Therefore, with this combination of spectral parameters, no solution to the CXB is possible. However, with $\langle $E$_{\rm cutoff} \rangle$ = 200 keV, a good fit to the CXB can be obtained.
In other words, some combinations of spectral parameters can be excluded because they do not reproduce the CXB for any XLF.

\begin{figure}[th]
	\centering
	\includegraphics[width=1.0\linewidth]{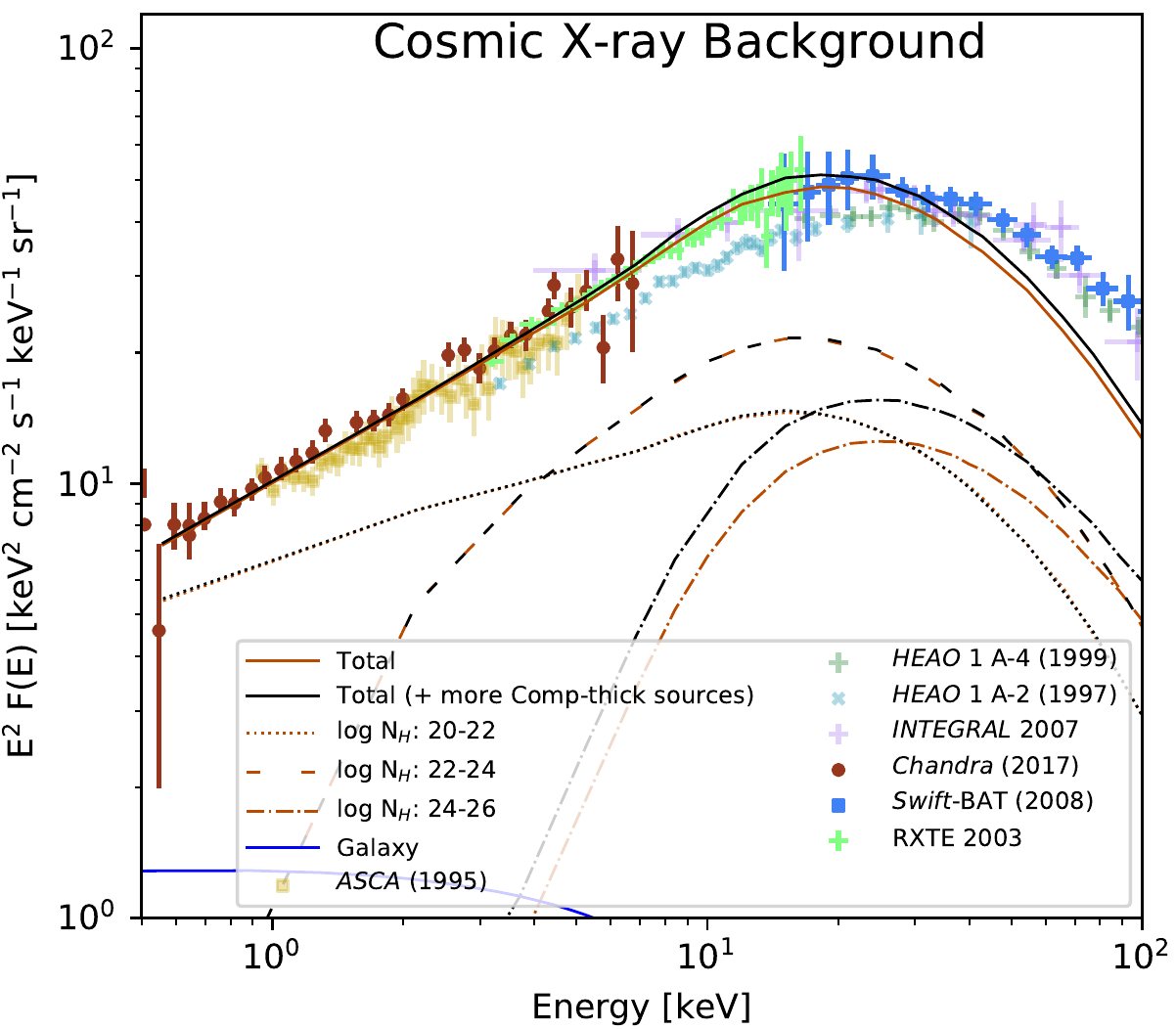}
	\caption{Cosmic X-ray background (CXB) with an unacceptably poor model fit. The data points are from \textit{Chandra} COSMOS (\citealp{nico2017}; \textit{red dots}), \textit{ASCA} (\citealp{ascasisxrb}; \textit{yellow squares}), \textit{RXTE} (\citealp{rxte}; \textit{bright green crosses}), \textit{Swift}-BAT (\citealp{ajello2008}; \textit{blue squares}), \textit{INTEGRAL} (\citealp{churazov2007}; \textit{pink crosses}), \textit{HEAO} 1 A-4 (\citealp{gruber1999hea0a4}; \textit{light green crosses}) and \textit{HEAO} 1 A-2 (\citealp{kinzer1997}; \textit{light blue crosses}).
	The \textit{HEAO} 1 data points are shown as example of data sets with discrepant normalizations relative to some of the most recent CXB measurements. 
	We do not include all CXB measurements in this figure, however with this example, we want to note that different instruments disagree in cross-normalization, and not the shape of the CXB. This issue, as well as the method of producing model fits is discussed in \S~\ref{ssec:neural_netowrk_method}. The model predictions ({\it black and brown lines}) were calculated using spectral parameters for Type~I objects in hard X-ray-observed Seyfert galaxies from \citet{derosa2012}; the \textit{brown lines} represent an X-ray luminosity function with lower space densities of heavily obscured Compton-thick objects, and the \textit{black lines} represent higher space densities of these objects. The different line styles represent CXB contributions from different absorption bins: unabsorbed ($\log$ (N$_{\rm H}/{\rm cm}^{\rm -2}<$ 22), Compton-thin ($\log$ N$_{\rm H}/{\rm cm}^{\rm -2}$ = 22$-$24) and Compton-thick ($\log$ N$_{\rm H}/{\rm cm}^{\rm -2} >$ 24) objects are shown separately with dotted, dashed and dash-dotted lines, respectively. 
	Each range of column density dominates in a different energy range of the CXB, with unabsorbed AGN dominating at low energies, Compton-thin objects at 3$-$10 keV, and Compton-thick objects adding significantly at E $= 10-60$ keV. In this example, unabsorbed and Compton-thin contributions fit the E $<$ 30 keV region well but at higher energies the CXB cannot be reproduced for  this set of spectral parameters. The galaxy contribution is calculated using the \citet{aird2015xlf} galaxy XLF but spectra with cutoff energy of 20-30~keV.} 
	\label{fig:derosa-cxb} 
\end{figure} 

In Paper~I, we illustrated this situation with several examples (see Figure~8 of that paper). Here, in Figure~\ref{fig:derosa-cxb} we show another example of reasonable (i.e., observed) X-ray spectra that nonetheless do not reproduce the CXB. 
Specifically, we assumed a set of spectral parameters derived from INTEGRAL hard X-ray observations (0.3-100 keV) of 33 Seyfert galaxies \citep{derosa2012}, as listed in Table~\ref{tab:spectral_parameters_in_papers} above. Figure~\ref{fig:derosa-cxb} shows the summed spectrum (solid lines) as well as the contribution from each column density bin (dashed lines). The unabsorbed 
contribution dominates at low energies (E $<$ 2 keV), the Compton-thin AGN contribution dominates at E $>$ 3 keV, and the Compton-thick contribution becomes significant at even higher energies (E $>$ 10 keV). This is consistent with the expectation that higher energy photons can escape even the thickest column densities. At low energies, the CXB effectively sets the number density of unabsorbed AGN, as there is negligible contribution from absorbed AGN.
At slightly higher energies, Compton-thin objects determine the slope of the CXB at 3$-$10 keV. 
In the example shown in Figure~\ref{fig:derosa-cxb}, the Compton-thin contribution cannot be increased any more without overproducing the CXB in this energy range; at the same time, this particular model undershoots the CXB at higher energies (E $>$ 60 keV), but increasing the Compton-thick contribution would overproduce CXB at lower energies.
This example illustrates how a particular set of spectral parameters --- even though observed in an X-ray survey --- cannot reproduce the observed CXB, and therefore cannot be representative of the full (underlying) ensemble of AGN.

Even without considering the CXB, we can see that parameter distributions for observed samples do not agree with each other (Table~\ref{tab:spectral_parameters_in_papers}). This may be caused by different spectral fitting procedures and/or by selection biases. So it is important to remember that parameters constrained in some survey samples may not be representative of the overall AGN population. Furthermore, they {\it cannot} be representative if they do not sum to the observed CXB.

Note that there are many degeneracies in fitting the CXB. For example, given a set of assumed spectra, the contribution in a given absorption bin can be produced by AGN of different luminosities (e.g., 100 AGN at $L_{\rm 2-10~keV} = 10^{42}$ erg s$^{\rm -1}$ are equivalent to one AGN at $L_{\rm 2-10~keV} = 10^{44}$ erg s$^{\rm -1}$).
In this work, we do not focus on producing a unique population synthesis model that can fit the CXB; instead, we identify distributions of spectral parameters that under no circumstances --- i.e., for no XLF --- are consistent with the CXB.
To do this, we use a neural network that fits the CXB by readjusting the XLF dependence on luminosity and/or column density. 

Absorption functions have been derived differently in previous works. For example, \citet{ueda2014} derived an X-ray luminosity function using a dozen X-ray surveys, then used the three surveys with the highest photon counts to formulate an absorption function over the $\log$ (N$_{\rm H}$/cm$^{-2}$) $= 20-24$ range. The Compton-thick fraction was assumed to be evenly spaced for $\log$(N$_{\rm H}$/cm$^{-2}$)= 24-26, with the same average number density as for $\log$(N$_{\rm H}$/cm$^{-2}$) =22-24. That is, the luminosity function and absorption function were derived separately. Other works derive both functions together. For example, \citet{buchner2015} used a sample of \textit{Chandra} hard-band-detected objects to produce an XLF that treats luminosity and absorption in equal footing using a Bayesian methodology, instead of evaluating the absorption function separately.

The neural network we introduced in Paper~I also varies the distribution of objects over luminosity and absorption bins simultaneously to converge on XLFs that can satisfy the CXB constraint. In this way, the absorption function is built into the luminosity function. As described in Paper~I, the neural network starts with the \citet{ueda2014} luminosity function, with the updated \textit{Swift}-BAT 70-month \citet{ricci2015} absorption function, and both functions are updated in each step of the neural network. As neural networks are sensitive to initial conditions, to ensure faster convergence, we start with 100 parallel networks, each of which re-weights the input XLF randomly. The weights associated with each luminosity and absorption bin are shown in Figure~6 and Appendix~B of Paper~I. The redshift dependence follows the observed distribution in \citet{treister2006}. 
If a fit to the CXB is possible for a spectral set, the neural network can converge on multiple XLFs that produce equally good fits, depending on the starting point of the network.

As discussed in Paper~I, there are dozens of other observed constraints besides the CXB that a correct XLF should be able to reproduce and that can help break the degeneracies and restrict the spectral parameter space further. For example, number counts, defined as the number of objects at a certain flux limit in a given band, are relatively insensitive to cutoff energy (for E $<$ 10 keV) yet are very sensitive to the luminosity distribution, which allowed us to break the degeneracy among XLFs and arrive at the solution presented in Paper~I. In the present work, however, we are not trying to produce a new population synthesis model --- we simply want to rule out sets of spectral parameters that can under no circumstances make up the CXB. Therefore, we only use CXB for the present analysis, ignoring the additional constraints considered in Paper~I (and avoiding their computational cost). 
Of course, tighter limits could be placed on spectral parameter space if those other constraints were incorporated, and we discuss this qualitatively in \S~\ref{sec:conclusion}.

\section{Method}\label{sec:method}

\subsection{Modeling AGN spectra}\label{ssec:spectra_method}

To model AGN spectra in this work, we use the sum of transmitted, reflected and scattered components, which has been shown to be the best way to model overall AGN X-ray spectra by \citet{buchner2014spectra}. Slight variations of this prescription were used by \citet{ueda2014}, \citet{buchner2015} and Paper~I. The \textsc{xspec} \citep{xspec} syntax of the spectral model is:

\textsc{fscatt $\times$ cutoffpl} + \textsc{tbabs $\times$ cabs $\times$ pexmon} + \textsc{borus02}.

\noindent
\textsc{borus02} \citep{mislav2018} is the most up-to-date, flexible prescription to model the reprocessing emission by a toroidal distribution of gas and dust. This model takes Compton-reflection by the torus into account, and prescribes an overall AGN spectrum 
similar to the above model, with \textsc{cutoffpl} instead of \textsc{pexmon}. As observed by \citet{dadina2008}, \citet{beckmann2009}, \citet{claudio2017bat} and \citet{zap2018}, unobscured objects usually have a higher 
reflection fraction than obscured objects, 
possibly due to reflection by the accretion disk (e.g., \citealp{beckmann2009}),
although the proportion of reflected to transmitted component for obscured objects is higher. This is why we include the \textsc{pexmon} model, which accounts for reflection due to a slab geometry like an accretion disk, which is not modeled as part of the conical geometry of \textsc{borus02}. We also include a reflection scaling factor of $\langle R \rangle$ $=$ 0 within our explored parameter space, which would correspond exactly to the model prescribed in \citet{mislav2018}. To account for  attenuation due to line-of-sight photoelectric absorption and Compton scattering, we multiply the reflected component by \textsc{tbabs} and \textsc{cabs}, respectively.

The free parameters in all these models are the photon index, $\Gamma$, cutoff energy, E$_{\rm cutoff}$, and normalization of power-law radiation from the corona; the reflection scaling factor, R; the inclination angle, $\theta_{\rm inc}$, and opening angle, $\theta_{\rm OA}$, of the torus; and the 
hydrogen column density, log(N$_{\rm H}$/cm$^{-2}$) along the line of sight. Absorbing column densities across all the \textsc{xspec} models (i.e., \textsc{tbabs}, \textsc{cabs}, \textsc{borus02}) are consistent. 
The distribution of AGN as a function of luminosity and absorbing column density is given by the XLF.

{\citet{marchesi2019} studied the torus covering factor of 35 Compton-thick objects using \textit{NuSTAR} spectra, finding an average covering factor $\simeq$ 0.5, i.e., $\theta_{\rm OA} \simeq 60^{\circ}$. For the \textit{Swift}-BAT 70-month AGN sample, \citet{claudio2017bat} found that for the 12 objects for which the opening angle 
of the torus could be constrained, the median value was similar, $\theta_{\rm OA}$ is $58 \pm 3^{\circ}$. Based on these results, we fix $\theta_{\rm OA} = 58^{\circ}$. For the \textsc{borus02} model, the line of sight column density is equal to the torus column density at all angles larger than the opening angle. To be physically consistent with the likelihood of drawing an inclination angle, when constructing \textsc{xspec} spectra, we draw from a distribution p($\theta_{\rm inc}$) $\propto \sin \theta_{\rm inc}$ in the range $\theta_{\rm inc} = 58^{\circ}-87^{\circ}$, i.e., between the opening angle of the torus and the highest value allowed by \textsc{borus02}. For each set of spectra, we use three Gaussian distributions for R, $\Gamma$ and $E_{\rm cutoff}$. We keep $\sigma_R$ and $\sigma_{\rm Ecut}$ constant for all distributions, equal to the values observed in the \textit{Swift}-BAT 70-months sample \citep{claudio2017bat}, which reports $\sigma_R = 0.09-0.14$ and $\sigma_{\rm Ecut} = 29-36$ keV. Each spectral set is defined by the mean values $\langle R \rangle$, $\langle \Gamma \rangle$ and $\langle  E_{\rm cutoff} \rangle$, and the dispersion in the photon index,  $\sigma_{\Gamma}$. The latter dispersion is included due to its strong influence on the CXB at high energies (see \S~\ref{sec:results_Bayesian}).

The reflection scaling factor is drawn from a Gaussian distribution with $\langle R \rangle$ in the range $0.0-2.0$, and a standard deviation $\sigma_R $ in the range $0-1.0$. For the \textsc{pexmon} model, both R and inclination angle affect the normalization of the reflection component; to avoid parameter coupling, we keep the inclination angle for this component fixed at the default value (60$^{\circ}$) and let R vary. The mean of the photon index, $\langle \Gamma \rangle$, is allowed to vary from 1.4 to 2.2, with a range of standard deviations: $\sigma_{\Gamma} = 0-0.5$. $\langle E_{\rm cutoff} \rangle$ varies from 50 to 500 keV, with a $\sigma_{\rm Ecut} = 0-100$ keV. 
The range spanned by each variable reflects the observed range reported in the literature (Table~\ref{tab:spectral_parameters_in_papers}). The example spectra shown in Figure~\ref{fig:gamma_variation_spectra} were all created using the \textsc{xspec} spectral model described here, and the parameters being varied are specified in each panel. Additionally, the opening angle of \textsc{borus02} is fixed at $\theta_{\rm OA} = 58^{\circ}$, and $\theta_{\rm inc} = 72^{\circ}$, as $\langle \theta_{\rm inc} \rangle \simeq 72^{\circ}$ for the p($\theta_{\rm inc}$) $\propto \sin \theta_{\rm inc}$ distribution. Note that these parameter distributions do not vary with redshift or luminosity in our model. However, these dependences can be incorporated in the future when better constraints become available.

As shown in Figure~\ref{fig:gamma_variation_spectra}, the absorbing column density significantly affects the shape of the spectra. As the absorption function is built into the luminosity function and determined by the neural network, we create a matrix of \textsc{xspec} AGN spectra with log (N$_{\rm H}$/cm$^{-2}$) at 60 evenly spaced points between 20$-$26. We use an algorithm called \hyperlink{https://pypi.org/project/vegas/}{\textsc{vegas}} \citep{vegas} to carry out the CXB integral; \textsc{vegas} performs Monte Carlo sampling over redshift, luminosity and absorption space $\simeq$ 10,000 times to calculate CXB at each photon energy. To sample the intensity for an AGN with log (N$_{\rm H}$/cm$^{-2}$) at an intermediate point between any two of these 60 evenly spaced spectra in absorption space, we use linear interpolation. This approach is more accurate than using a single reference log (N$_{\rm H}$/cm$^{-2}$) spectra for each absorption bin, as the spectra can vary considerably over a dex of log (N$_{\rm H}$/cm$^{-2}$), and more computationally efficient than producing the exact \textsc{xspec} spectra for each Monte Carlo sampling point while performing the integral.

\subsection{MCMC Sampling of Parameter Space Assuming an XLF}\label{sec:Bayesian_method}

 We first investigate the parameter space that reproduces the CXB assuming that the XLF we presented in Paper~I is representative of the underlying AGN population. We use a Bayesian approach to calculate the goodness of fit. Specifically, we use observed parameter spaces from \citet{claudio2017bat} to define the prior probability distributions and a Monte Carlo Markov Chain (MCMC) sampler to find the Gaussian log likelihood using observed CXB data points. 
 
 
 The MCMC sampler uses ensemble sampling: it deploys 250 walkers to start from a random point in the parameter space, uniformly over $\langle \Gamma \rangle = 1.4-2.2$, $\langle E_{\rm cutoff} \rangle = 50-500$ keV and $\langle R \rangle = 0.0-2.0$. We explore how the parameter spaces for the other three parameters shift assuming $\sigma_{\Gamma} = 0.1, 0.2, 0.3, 0.4, 0.5$. For this analysis, we fix $\sigma_E$ and $\sigma_R$ at the values observed for the \citet{claudio2017bat} unabsorbed AGN sample, namely, 36 keV and 0.14, respectively.
 
\subsection{XLF-Independent Exploration of Parameter Space using a Neural Network}\label{ssec:neural_netowrk_method}

The second approach in our investigation is more conservative because it makes no assumptions about the correct XLF. 
Instead, we assume a set of spectra and forward model to find a good fit to the CXB by varying the underlying XLF. As demonstrated earlier, some spectral sets do not produce good fits to the CXB regardless of the underlying AGN XLF; these can be ruled out definitively. We vary the underlying XLF in an automated way using the neural network we introduced in Paper~I
(described in detail in Figure~6 and Appendix~B of that paper). 
Given an input spectrum, the neural network adjusts a preliminary estimate of space densities (the \citealp{ueda2014} XLF with a modified absorption function from \citealp{ricci2015}) in luminosity and absorption bins to find the best possible fit to the CXB. The neural network converges on an XLF by attempting to reduce the difference between the observed CXB and the model predictions. The fit to the CXB only determines that correct proportions in absorption bins are possible for this spectral set. To produce the full population synthesis model described in Paper~I, further validation checks were made to rule out discontinuous solutions that did not reproduce the number counts 
or other constraints. Here, due to computational expense, we simply identify spectra that do not produce the CXB under any circumstances; this means that some of the ``allowed" parameter space may not in fact be viable (because it cannot match the untested constraints), but the excluded parameter space is definitely not viable. 

The parameter space explored in this MCMC approach is the same as in \S~\ref{sec:Bayesian_method}. Once the neural network has converged on the best possible fit to the CXB for each of the underlying spectra, we quantitatively determine the goodness of fit using the $\chi^2$ statistic.

For both the approaches described above, we use the CXB observed data points from \textit{Chandra} COSMOS \citep{nico2017}, \textit{RXTE} \citep{rxte} and \textit{Swift}-BAT \citep{ajello2008}, as these are the latest and most sensitive measurements, and have consistent normalization in overlapping regions. 
At low energies, \textit{ASCA} \citep{ascasisxrb} and \textit{Chandra} agree within the errors of those data sets, as shown in Figure~\ref{fig:derosa-cxb}. 
All recent measurements disagree with the older \textit{HEAO} 1 data (\citealp{gruber1999hea0a4,kinzer1997}), as discussed previously in the literature \citep{treister2009}. We fit the data considering two possible sets of errors: (A) uncertainties published for each individual CXB data set, shown by error bars in Figure~\ref{fig:derosa-cxb}, or (B) the larger empirical differences across all CXB data (i.e., incorporating the historically mismatched normalizations).


We use the $\chi^2$ statistic to judge the quality of model fits to each of the observed CXB data sets. We calculate the $\chi^2$ statistic as follows:

\begin{equation}
\chi^2 = \sum_i \frac{({\rm Observed}_i - {\rm Expected}_i)^2}{\sigma_i^2} .
\end{equation}

\noindent
Here, $\sigma_i$ is the measurement error given by either choice (A) or (B) above, yielding values $\chi_A^2$ or $\chi_B^2$ and associated probabilities. We found that 
choice (B) leads to formally acceptable but clearly poor fits to the shape of the CXB (see Figure~\ref{fig:derosa-cxb}), so we adopt choice (A) for the rest of the analysis. This point is discussed further in \S~\ref{sec:conclusion}. 


As for the Bayesian analysis, we fixed $\sigma_E=36$ keV and $\sigma_R=0.14$, corresponding to the observed values for unobscured objects \citep{claudio2017bat}. We provide an interactive online tool that allows users to explore the dependence of all parameters on absorption, 
as well as different values of $\sigma_E$ and $\sigma_R$. This is because many of the AGN samples in Table~\ref{tab:spectral_parameters_in_papers} report significantly different parameters for different ranges of absorbing column density. We do not incorporate a possible luminosity and redshift dependence in this work, as discussed further in \S~\ref{sec:conclusion}.

\section{Results}\label{sec:results}

\subsection{Bayesian Analysis with Fixed XLF}\label{sec:results_Bayesian}

As shown in the trace plots in Figure~\ref{fig:convergence}, the ensemble sampler for the Bayesian approach converges in fewer than 3000 steps. The results of the Bayesian analysis, summarized in Table~\ref{tab:Bayesian_1sig_2sig_distribution}, show how the mean parameter values shift as the dispersion in photon index is changed: $\langle E_{\rm cutoff} \rangle$ and $\langle R \rangle$ decrease with increasing $\sigma_\Gamma$, while $\langle \Gamma \rangle$ increases. Increasing $\sigma_{\Gamma}$ adds photons at high energies, so $\langle E_{\rm cutoff} \rangle$, $\langle R \rangle$, and $\langle \Gamma \rangle$ shift to decrease high-energy photons. Figure~\ref{fig:all_1D_hists} shows the shift in 1D distributions with $\sigma_\Gamma$ for each parameter. The reflection parameter is least affected, moving to slightly lower values as $\sigma_\Gamma$ increases. The most significant dependence is on the cutoff energy, for which the distribution narrows significantly and moves to lower energies as $\sigma_\Gamma$ increases. $\langle \Gamma \rangle$ shifts to higher values with increasing $\sigma_\Gamma$. As increasing $\sigma_\Gamma$ adds more photons at higher energies, all these distributions shift to accommodate the increase in photons in that region.
The 2D distributions show that for any $\sigma_\Gamma$, the means of all three parameters are positively correlated with each other. 

\begin{deluxetable}{lllrll}[th]
\tablewidth{0pt}
\tablecaption{\label{tab:Bayesian_1sig_2sig_distribution}\textsc{Spectral parameter distributions for Bayesian analysis with fixed XLF.}} 
 \tablehead{\colhead{\textsc{$\Gamma$ dispersion}} & \colhead{\textsc{$\langle R \rangle$}} & \colhead{\textsc{$\langle E_{\rm cutoff} \rangle$}} & \colhead{\textsc{$\langle \Gamma \rangle$}}}
\startdata
                \hline$\sigma_{\Gamma} = 0.1$ & $1.071^{+0.097}_{-0.091}$ & $204^{+36}_{-20}$ & $1.984\pm 0.020$  \\ 
		$\sigma_{\Gamma} = 0.2$ & $1.063^{+0.088}_{-0.097}$ & $180^{+14}_{-23}$ & $1.996^{+0.030}_{-0.020}$  \\ 
		$\sigma_{\Gamma} = 0.3$ & $0.999^{+0.093}_{-0.094}$ & $136\pm 12$ & $2.019^{+0.035}_{-0.029}$  \\ 
		$\sigma_{\Gamma} = 0.4$ & $0.941^{+0.097}_{-0.093}$ & $113.4^{+10.0}_{-9.1}$ & $2.049^{+0.034}_{-0.037}$ \\ 
		$\sigma_{\Gamma} = 0.5$ & $0.902^{+0.097}_{-0.096}$ & $101.0^{+9.5}_{-8.7}$ & $2.083^{+0.040}_{-0.044}$ \\ 
\enddata
\end{deluxetable}

\begin{figure}[th]
	\centering
	\includegraphics[width=1.0\linewidth]{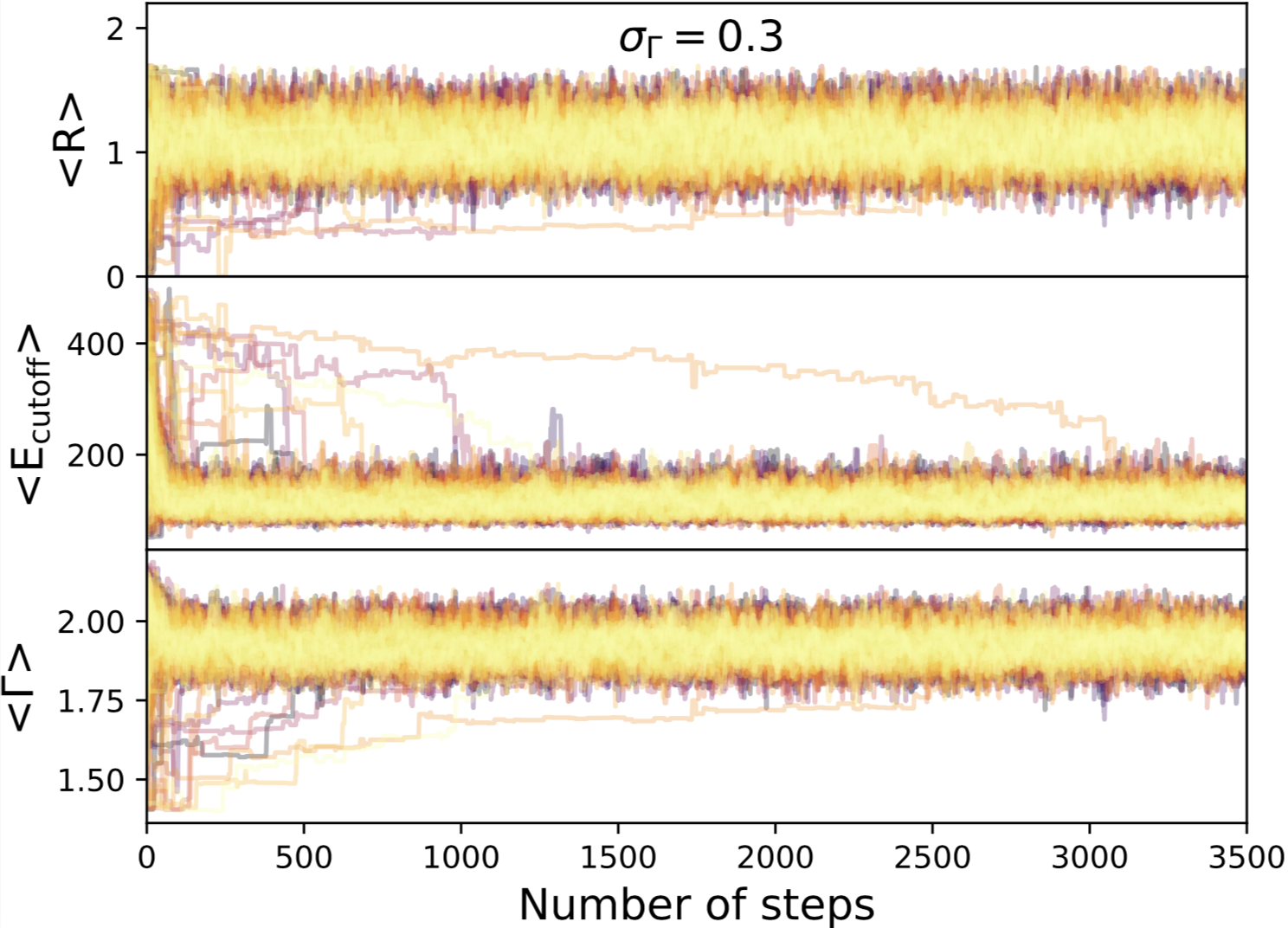}
	\caption{An example of convergence of the ensemble sampler in the Bayesian analysis part for the $\sigma_\Gamma = 0.3$ case. Each line represents one `walker' of the ensemble sampler, which explores the parameter space to find regions that best reproduce the observed CXB.} 
	\label{fig:convergence} 
\end{figure} 

\begin{figure}[th]
	\centering
	\includegraphics[width=1\linewidth]{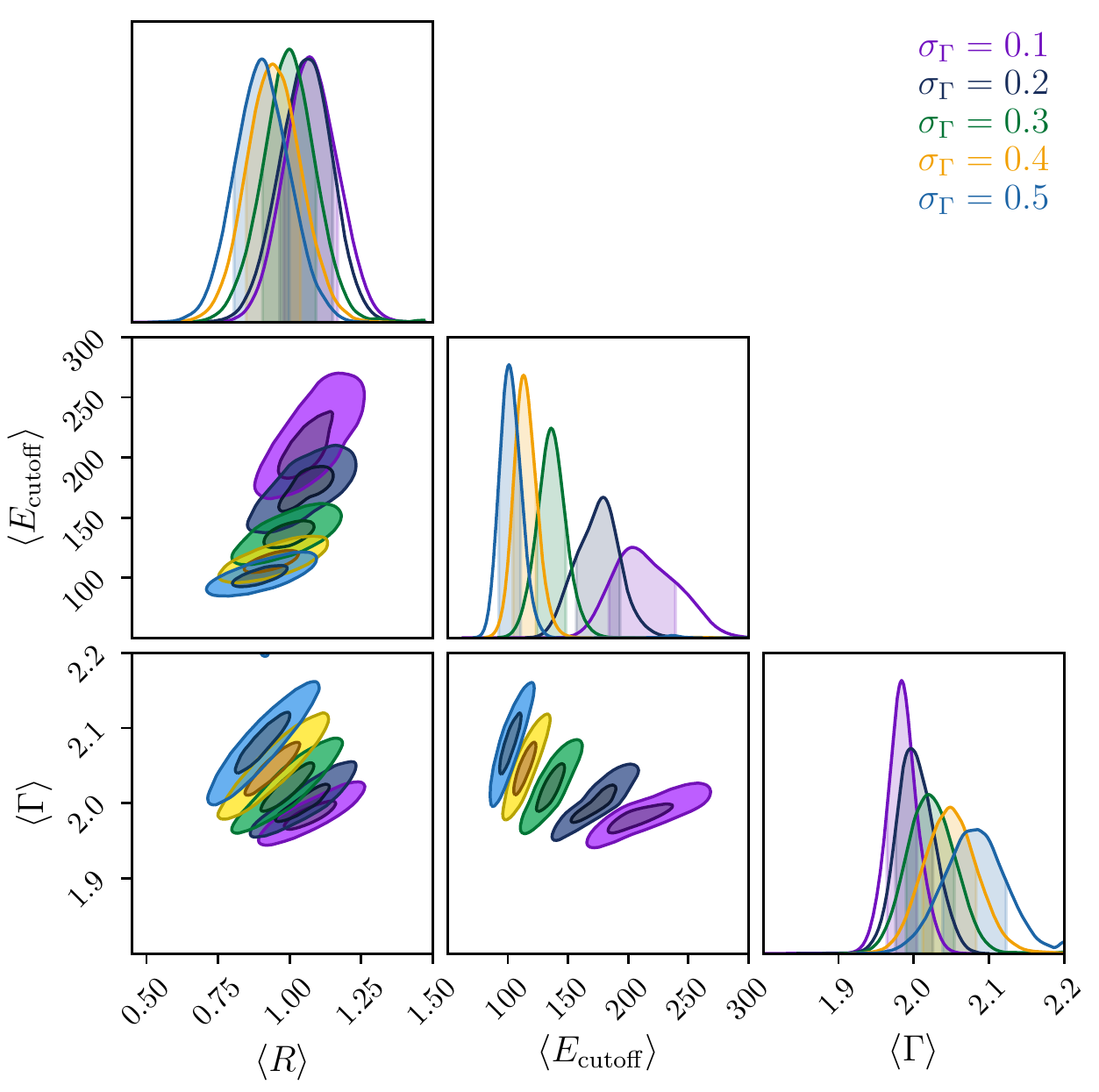}~
	\caption{Parameter distributions allowed by the CXB constraint and the XLF presented in \citet{ananna2018}. Each color represents the result for a spectral set assuming a fixed dispersion in photon index, $\sigma_{\Gamma} = 0.1, 0.2, 0.3, 0.4$ or $0.5$. Diagonally, the one-dimensional histograms show distribution of each parameter. The contour plots show 1$\sigma$ and 2$\sigma$ contour levels in 2D for each pair of parameters. This figure shows that for a given XLF, multiple spectral sets can produce fits to the CXB because the parameters are degenerate.} 
	\label{fig:all_1D_hists} 
\end{figure} 

\begin{figure}[th]
	\centering
	\includegraphics[width=1.\linewidth]{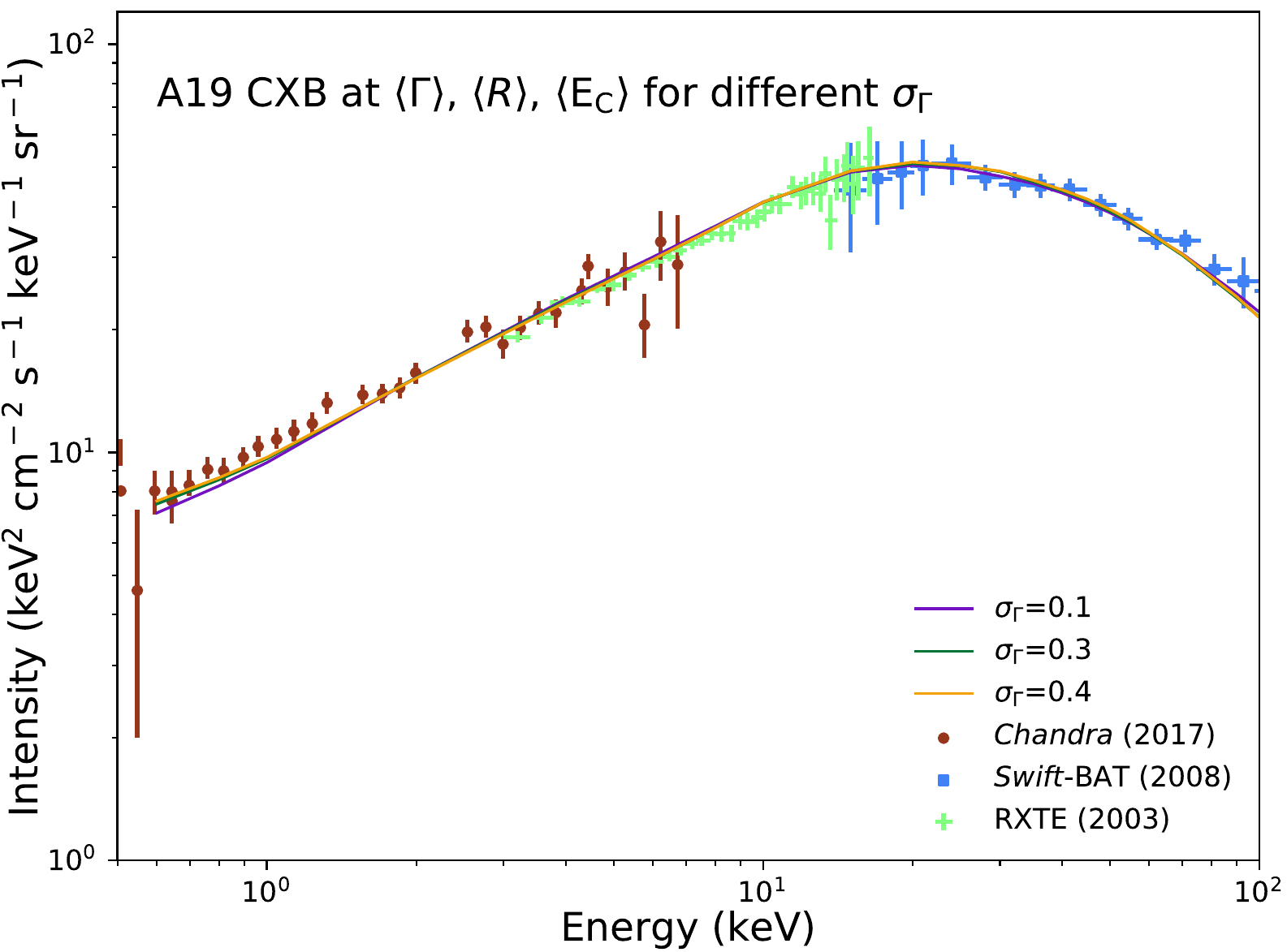}
	\caption{Fits to the CXB obtained for the XLF presented in \citet{ananna2018} and three examples of the spectral parameters defined in Table~\ref{tab:Bayesian_1sig_2sig_distribution} ($\sigma_\Gamma = 0.1,0.3$ and 0.4. The different models ({\it solid lines}) determined by the Bayesian analysis provide equally good fits, illustrating the degeneracy of the parameters. 
	The data points are from \textit{Chandra} COSMOS (\citealp{nico2017}; \textit{red dots}), \textit{RXTE} (\citealp{rxte}; \textit{bright green crosses}) and \textit{Swift}-BAT (\citealp{ajello2008}).}
	\label{fig:cxb_five_lines} 
\end{figure} 

Figure~\ref{fig:cxb_five_lines} shows that using the parameter combinations in each row of Table~\ref{tab:Bayesian_1sig_2sig_distribution}, a perfect fit to the CXB can be produced using the same XLF, demonstrating the strong degeneracies among these parameters --- in effect, the low cutoff energy cancels out the contribution at high energies of a large dispersion in photon index.

We note that starting with different priors, such as the \textit{Swift}-BAT 70-month observed parameter distribution from \cite{claudio2017bat}, our results still converge on the same posterior distributions.

\subsection{Neural Network Fitting with Arbitrary XLFs}\label{ssec:neural_network_results}

\begin{figure*}[th]
	\centering
	\includegraphics[width=0.8\linewidth]{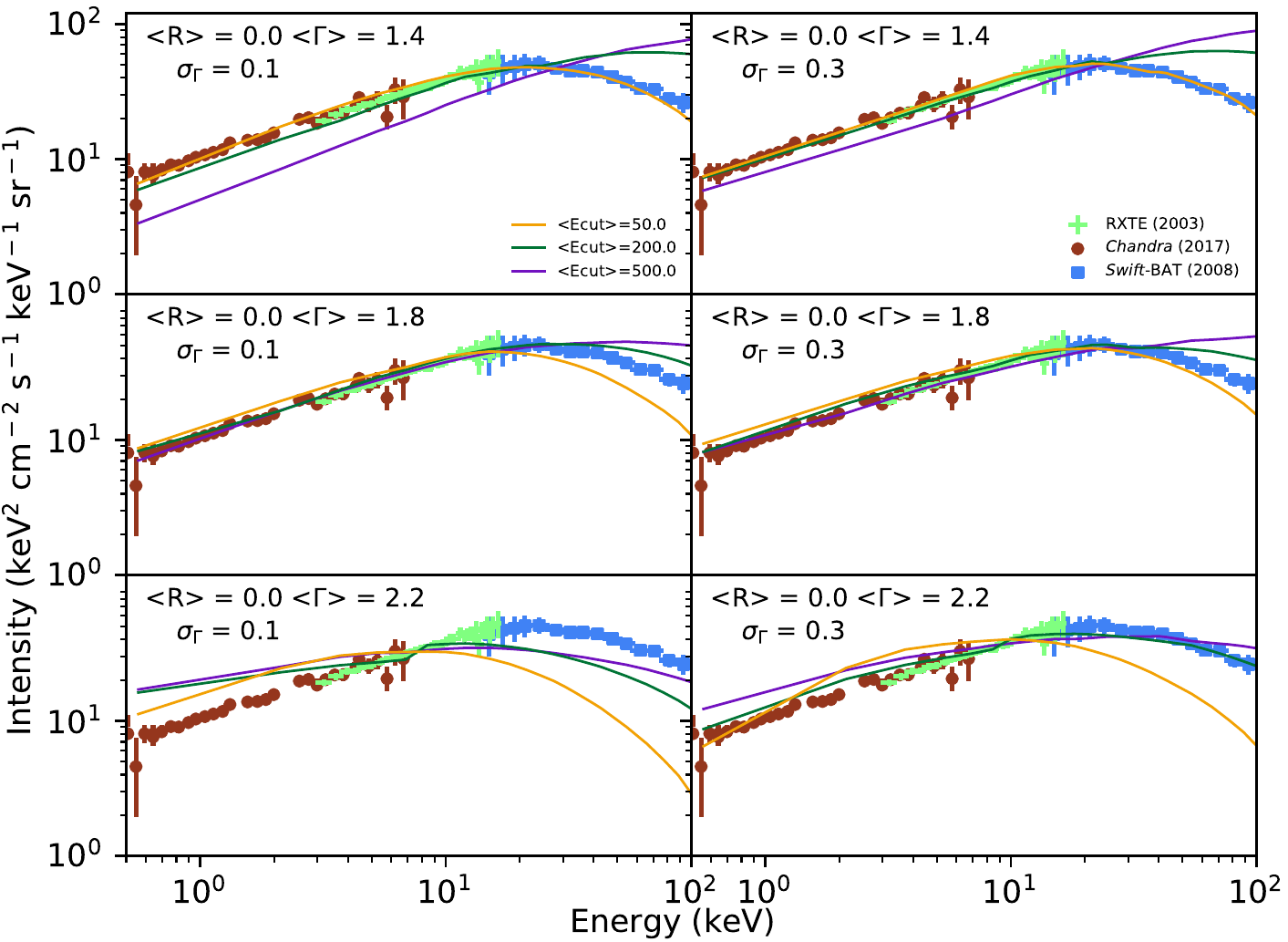}
	\includegraphics[width=0.8\linewidth]{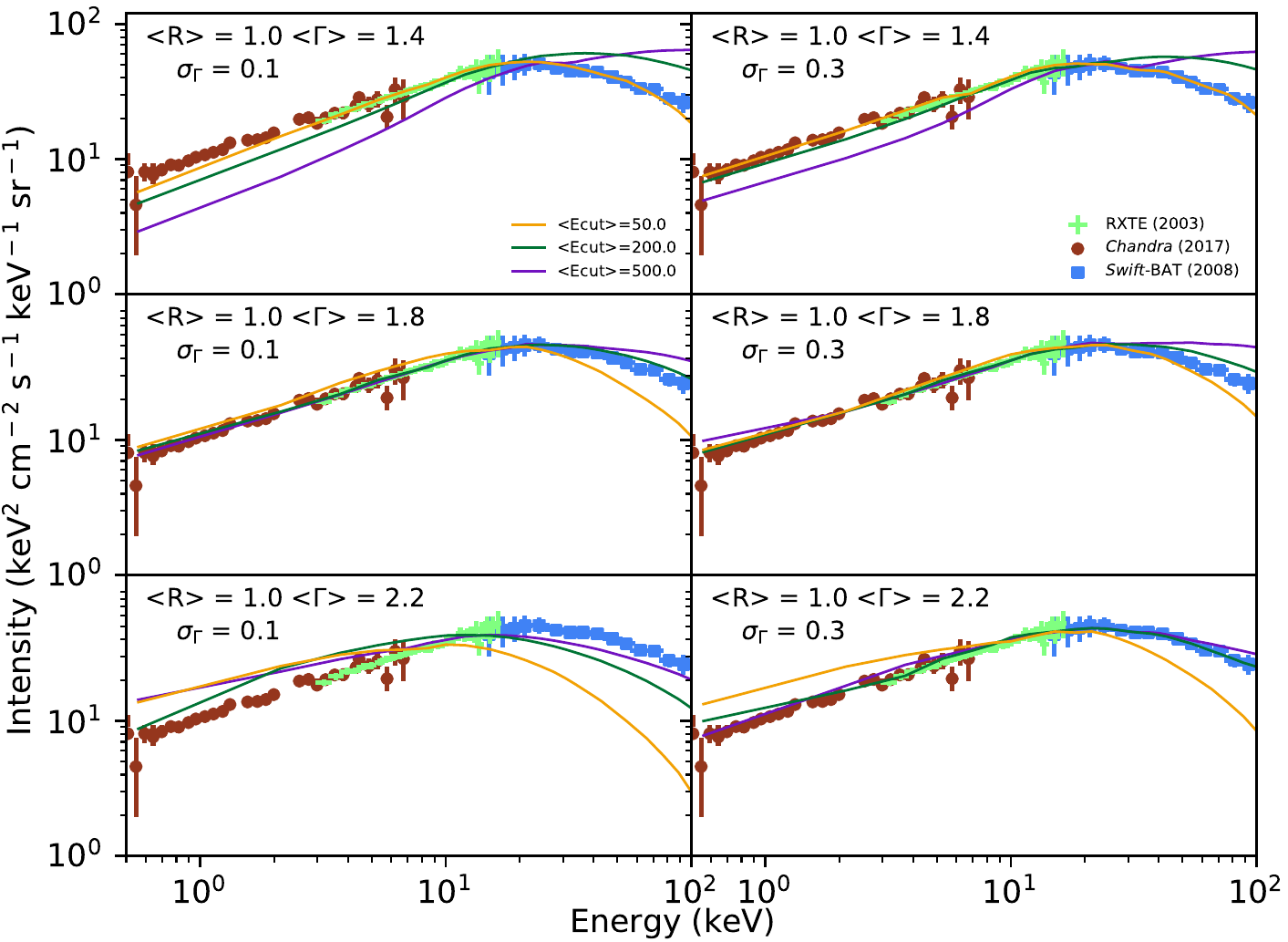}
	\caption{Examples of the best possible --- and mostly unacceptable --- fits to the CXB data for each of twelve different sets of spectral parameters. The top six panels show fits for $\langle R \rangle$ = 0, and the bottom six panels show fits for $\langle R \rangle$ = 1.0. Each row corresponds to a different photon index: 
	$\langle \Gamma \rangle$= 1.4
	(\textit{first and fourth rows}), 
	$\langle \Gamma \rangle = 1.8$ 
	(\textit{second and fifth rows}), 
	and $\langle \Gamma \rangle = 2.2$ 
	(\textit{third and sixth rows}).
	The \textit{left panels} show results for $\sigma_{\rm \Gamma} = 0.1$ and \textit{right panels} for $\sigma_{\rm \Gamma} = 0.3$. 
	The standard deviations for cutoff energy and R are fixed at $\sigma_{\rm Ecut} = 36$ keV and $\sigma_{\rm R} = 0.14$. The \textit{solid yellow lines}, \textit{solid green lines} and \textit{solid purple lines} shows the CXB predictions for $\langle$E$_{\rm cutoff} \rangle$ = 50, 200 and 500 keV, respectively. The data points are as in Figure~\ref{fig:cxb_five_lines}.} 
	\label{fig:xrb_neural_network} 
\end{figure*} 

\begin{figure*}[th]
	\centering
	\includegraphics[width=1.\linewidth]{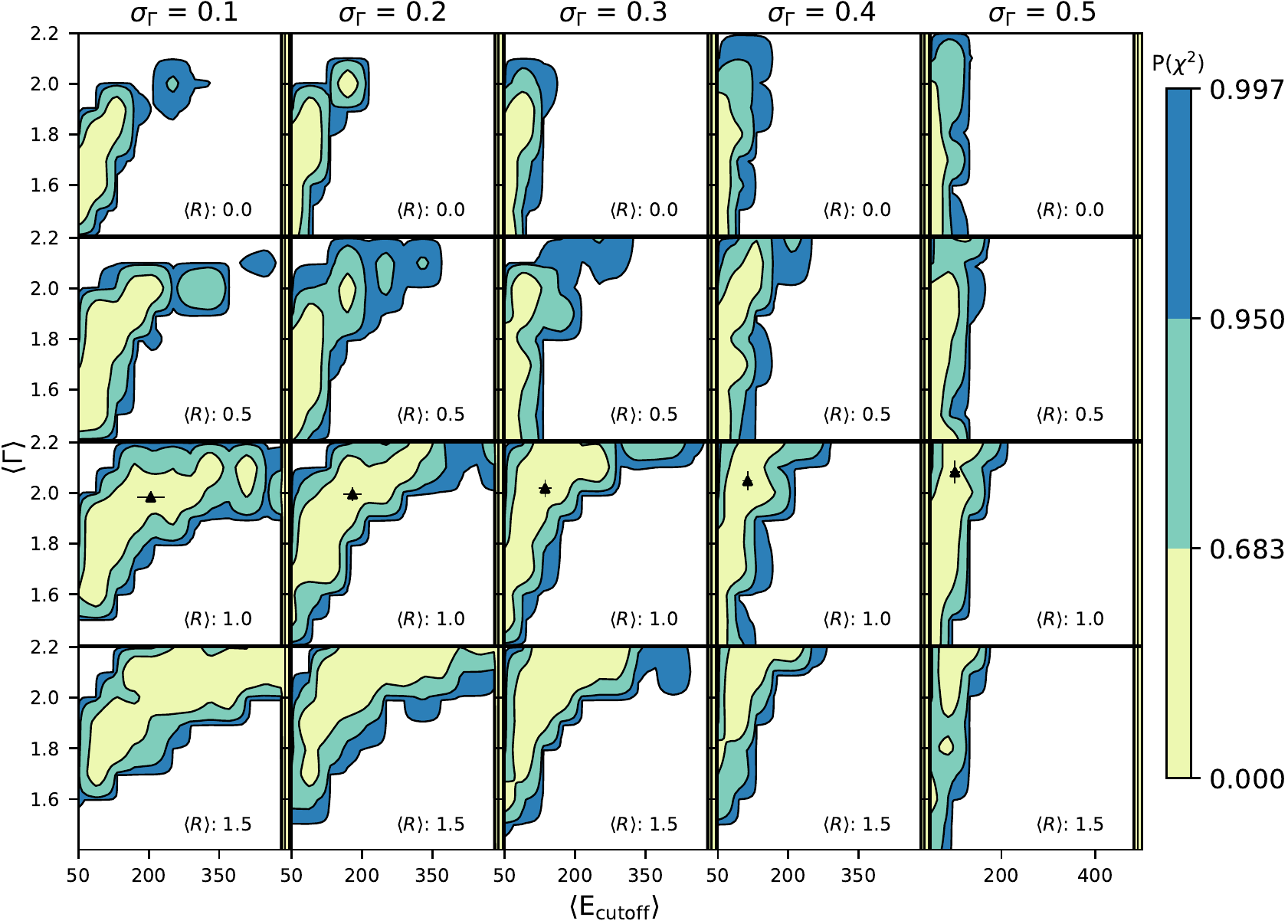}
	\caption{Allowed regions of the spectral parameter space, for unconstrained XLF. Contours for 1, 2 and 3$\sigma$ are shown in yellow, green and blue, respectively.
	The black triangular data points indicate results of the Bayesian analysis using the XLF from Paper~I. To create this figure, the same spectral set was assumed for all absorption bins. The values of  $\sigma_E$ and $\sigma_R$ were kept constant at 36 keV and 0.14, respectively. The entire probability space can be explored quantitatively using the 
	interactive tool described in \S~\ref{ssec:neural_network_results}, 
	which allows users to specify a different spectral set for each absorption bin and explore other values of $\sigma_E$ and $\sigma_R$ as well as the variables shown in this plot.}
	\label{fig:neural_network_result} 
\end{figure*} 

The neural network finds the underlying AGN population that, given an input set of spectral parameters, best fits the CXB. This allows us to rule out spectral parameters for which a good fit to the CXB is not possible under any circumstances.

Figure~\ref{fig:xrb_neural_network} qualitatively demonstrates the interplay of the parameters $\langle \Gamma \rangle$ and $\langle E_{\rm cutoff} \rangle$ in fitting the CXB for snapshots of the parameter space. 
Specifically, each panel shows examples of best fits to the CXB that were generated by the neural network by varying $\langle E_{\rm cutoff} \rangle$ between 50$-$500 keV while keeping $\langle \Gamma \rangle$ and $\langle R \rangle$ constant. As $\langle \Gamma \rangle$ increases, the best-fit CXB model shifts upwards at lower photon energies and downwards at higher energies. 
With photon index as low as $\langle \Gamma \rangle = 1.4$, a very low cutoff energy ($\langle E_{\rm cutoff} \rangle \sim 50$~keV) fits the CXB consistently at both high and low energies. As $\langle \Gamma \rangle$ steepens to 1.8, 
the low cutoff energy causes the CXB to drop too quickly at high energies, especially with a low dispersion, $\sigma_{\Gamma}=0.1$. 
This effect is more pronounced with an even steeper spectrum, e.g.,  $\langle \Gamma \rangle = 2.2$. From the middle panels, where $\langle \Gamma \rangle = 1.8$, it is clear that a good fit to the CXB can be obtained with $\langle E_{\rm cutoff} \rangle$ close to but smaller than 200 keV; for $\langle \Gamma \rangle = 2.2$, no solutions are possible, at least when $\langle R \rangle = 0$ (upper three panels). 

A higher photon dispersion (right panels) changes little at low photon energies but adds photons at higher energies, which causes the best-fit $\langle E_{\rm cutoff} \rangle$ to shift lower. This effect is more conspicuous in the middle and lower panels, for both values of $\langle R \rangle$. Note that in this work, the value of $\langle R \rangle$ represents reflection from the accretion disk rather than the torus, and this component contributes significantly only at intermediate energies, as shown in Figure~\ref{fig:gamma_variation_spectra}, in the same region where heavily obscured objects contribute the most. When a higher $\langle R \rangle$ value is assumed, the XLF readjusts by reducing the space densities of Compton-thick AGN. As shown in Figure~\ref{fig:xrb_neural_network}, the fit improves at intermediate energies when a non-zero $\langle R \rangle$ is assumed (lower three panels).

We quantitatively determined and report the quality of model fits (such as the lines shown in Figure~\ref{fig:xrb_neural_network}) to the observed CXB data points, 
for the entire range of parameter space described in \S~\ref{sec:Bayesian_method}. The shape of the probability distribution as a result of the XLF-independent analysis is complex, as shown in Figure~\ref{fig:neural_network_result}.  This figure shows the $\chi^2$ probability distribution up to 3$\sigma$ for snapshots of the parameter space assuming the same parameter distribution for all absorption bins, and fixed  dispersions $\sigma_R = 0.14$ and $\sigma_E = 36$ keV. We provide an interactive tool to determine the exact probability at any point in this parameter space, at \url{https://colab.research.google.com/drive/1eBA9gZX3yiTLxqhD7CjiLJ9hV6er56W3}. This tool also allows users to run the neural network using different parameter distributions on different absorption bins, and for other values of $\sigma_R$ and $\sigma_E$. We used MCMC sampling on the results of the neural network to find the allowed distribution for each parameter (marginalizing over the others): $\langle R \rangle = 0.99^{+0.11}_{-0.26}$, $\langle E_{\rm cutoff} \rangle = 118^{+24}_{-23}$, $\sigma_\Gamma = 0.101^{+0.097}_{-0.001}$ and $\langle \Gamma \rangle = 1.9^{+0.08}_{-0.09}$.} 

Figure~\ref{fig:neural_network_result} shows that as $\sigma_\Gamma$ increases, the allowed region shifts to smaller values of cutoff energy; however, the shift in $\langle \Gamma \rangle$ with $\sigma_\Gamma$ is not as significant. As $\langle R \rangle$ is increased, the allowed region shifts
to higher values of $\langle \Gamma \rangle$. This figure demonstrates that the CXB can only be produced for a continuous swath of the parameter space, and the intrinsic AGN spectra must lie in this region. 


In Table~\ref{tab:sample_results}, we show the goodness of fit to the CXB assuming parameter spaces of surveys described in Table~\ref{tab:spectral_parameters_in_papers}. We accept a part of the parameter space if it can fit the CXB within 3$\sigma$ significance level. These results are discussed in \S~\ref{sec:conclusion}.

\begin{deluxetable*}{llll}[th]
\tablewidth{0pt}
\tablecaption{\label{tab:sample_results}\textsc{Probability that published parameters for AGN samples from Table 1 fit the CXB\tablenotemark{1}}}
 \tablehead{\colhead{\textsc{Sample}} & \colhead{\textsc{Selection}} & \colhead{\textsc{P($\chi^2_{\rm A}$)~~\tablenotemark{2}}} &
 \colhead{\textsc{Result}}}
\startdata
\citet{claudio2017bat} & $14-195$ keV & $2-3\sigma$ & Acceptable \tablenotemark{3} \\
\citet{zap2018}\tablenotemark{4} & $8-24$ keV & $2-3\sigma$ &  Acceptable  \\
\citet{malizia2014}\tablenotemark{5} & $0.3-100$ keV selected Type-1 & $3-4\sigma$ & Rejected \\
\citet{buchner2014spectra}\tablenotemark{6} & $0.5-8$ keV & $> 5\sigma$ & Rejected \\
\citet{bntorus2011}\tablenotemark{6} & IR selected  & $> 5\sigma$ & Rejected \\
\citet{scott2011}\tablenotemark{6} & Optically selected Type-1 & $> 5\sigma$ & Rejected  \\
\citet{beckmann2009}\tablenotemark{7} & $> 20$ keV & $<1\sigma$ & Acceptable \\
\citet{dadina2008} & $2-100$ keV local & $4-5\sigma$ & Rejected \\
\citet{derosa2012}\tablenotemark{8} &  $>20$ keV & $1-2\sigma$ & Acceptable \\
\citet{ueda2014} & $14-195$ keV (9 month sample) & $> 5\sigma$ & Rejected \\
\enddata
\tablenotetext{1}{  The significance levels and probabilities for any region of the parameter space can be quantified by running the neural network using an interactive tool (link in \S~\ref{ssec:neural_network_results}).}
\tablenotetext{2}{ Probability range with respect to the errors associated with individual data sets (shown using error bars in Figure~\ref{fig:derosa-cxb}).}
\tablenotetext{3}{ We used the parameter distributions shown in Table~\ref{tab:spectral_parameters_in_papers}, with three different sets of parameters for unabsorbed, Compton-thin and Compton-thick objects, except as noted below in footnotes 5-9.} 
\tablenotetext{4}{ Assigned $\sigma_R = 1.0$ and $\sigma_R = 0.5$ for the unabsorbed and the absorbed sources, respectively, based on the 25th and 75th percentile values of R.}
\tablenotetext{5}{ Assuming no reflection scaling factor.}
\tablenotetext{6}{ As no cutoff energy is reported, we set the value to 500 keV.
For \citet{bntorus2011} we also assume no reflection scaling factor.}
\tablenotetext{7}{ The exact prescription used for unabsorbed sources: $\langle R \rangle=1.2$, $\sigma_R =0.45$, $\langle \Gamma \rangle =  1.96$, $\sigma_\Gamma = 0.02$, $\langle E_{\rm cutoff} \rangle = 86.0$, $\sigma_E = 17.0$. For absorbed sources: $\langle R \rangle=1.1$, $\sigma_R =0.59$, $\langle \Gamma \rangle =  1.91$, $\sigma_\Gamma = 0.02$, $\langle E_{\rm cutoff} \rangle = 376.0$, $\sigma_E = 42.0$.}
\tablenotetext{8}{ A dispersion of $\sigma_R=1.0$ is applied here, which is the highest allowed value in our model. Note that the `Acceptable' parameter space has different parameter distributions for absorbed and unabsorbed AGN. Figure~\ref{fig:derosa-cxb} shows an example of a poor fit by assuming the parameter space of unabsorbed AGN for all AGN.}
\end{deluxetable*}

\section{Discussion and Conclusions}\label{sec:conclusion}

In Paper~I, we reported that some previous XLFs do not reproduce the CXB with the AGN spectral sets reported in those works, and found that the CXB can only be reproduced within certain parameter ranges. In this paper, as well as in Paper~I, we show that some combinations of spectral parameters --- even those that are observed in X-ray surveys --- cannot reproduce the CXB for any underlying XLF. This is possibly due to selection biases in survey samples, such that those observed spectra are not representative, as well as the fact that constraining certain spectral parameters is possible only for objects with the highest photon counts, further biasing the measurements. 

In this work, we tried a different approach to explore the parameter space of AGN: we assumed a set of spectra and forward modeled to fit the CXB, first with the most recent, best constrained XLF, then by allowing the XLF to vary arbitrarily. While parameter spaces that can reproduce the CXB may not be representative of the real AGN population, we can definitely rule out regions of parameter space that do not satisfy the CXB constraint. For example, for the fixed dispersions $\sigma_E = 36$ keV and $\sigma_R = 0.14$, a spectral set with $\langle \Gamma \rangle = 1.7, \sigma_\Gamma = 0.2$, $\langle E_{\rm cutoff} \rangle = 300$ keV is not able to reproduce the CXB because both $\sigma_\Gamma$ and the high cutoff energy add too many photons at the high energy end. In such cases, lowering the average cutoff energy to around 200 keV, or lowering $\sigma_\Gamma$, or increasing the $\langle \Gamma \rangle$ reduces the number of photons at high energies, and then an acceptable fit to the CXB can be produced.

In \S~\ref{sec:results_Bayesian} we discussed the distribution of spectral parameters that can produce the CXB assuming the most up-to-date X-ray luminosity function (from Paper~I). Figure~\ref{fig:all_1D_hists} illustrates that $\langle \Gamma \rangle$, $\langle R \rangle$ and $\langle$E$_{\rm cutoff} \rangle$ are completely degenerate. The change in $\langle \Gamma \rangle$, $\langle R \rangle$ and $\langle$ E$_{\rm cutoff} \rangle$ in response to increasing $\sigma_\Gamma$ always decreases photons at higher energies, to compensate for a higher $\sigma_\Gamma$. This is because a higher $\sigma_\Gamma$ effectively adds more photons in that region.

All five assumed $\sigma_\Gamma$ values reproduce the CXB perfectly with the XLF from Paper~I; however, changing $\sigma_\Gamma$ changes the other three parameters significantly. This demonstrates that multiple sets of spectra can satisfy the CXB when coupled with a given XLF. As shown in Figure~\ref{fig:all_1D_hists}, the most noticeable effect of increasing $\sigma_\Gamma$ occurs to $\langle$E$_{\rm cutoff} \rangle$; this is especially apparent in the one-dimensional histograms of $\langle$E$_{\rm cutoff} \rangle$, which have the least overlap. This is expected as $\langle$E$_{\rm cutoff} \rangle$ has to decrease with increasing $\sigma_\Gamma$ in order to prevent over-producing the high energy end of the CXB. The significant change indicates that $\langle$E$_{\rm cutoff} \rangle$ more directly contributes to decreasing photons at higher energies than the other two parameters.


In Figure~\ref{fig:neural_network_result}, we show results for part of the parameter space for the XLF-independent approach using the neural network. This figure shows that if we keep dispersions for all the distributions constant, increasing $\langle R \rangle$ shifts the allowed region to higher values of $\langle \Gamma \rangle$ and $\langle$E$_{\rm cutoff} \rangle$. For non-zero $\langle R \rangle$, increasing $\sigma_\Gamma$ shifts the allowed region to lower cutoff energies, which is consistent with the results from \S~\ref{sec:results_Bayesian} --- the high $\sigma_\Gamma$ needs to be offset by low $\langle$E$_{\rm cutoff} \rangle$ and high $\langle \Gamma \rangle$ to prevent overproduction of the CXB at high energies.


Increasing $\langle R \rangle$ at a fixed $\sigma_\Gamma$ also expands the allowed region upwards towards higher $\langle \Gamma \rangle$, and shifts it rightwards towards higher $\langle$E$_{\rm cutoff} \rangle$. Conversely, when the underlying AGN spectra are relatively flat, it is difficult to reproduce the drop in intensity of the CXB at E $> 40$ keV, unless the cutoff energy is very low. Furthermore, if a high $\langle R \rangle$ is assumed, the `Compton-hump' region of the CXB is dominated by the reflected contribution, which then drops at higher photon energies. As a result, increasing $\langle R \rangle$ expands the allowed region to higher cutoff energies.
 
In Table~\ref{tab:sample_results}, we show the results of the neural network analysis for each observed data set. 
Choice (B) errors allow a broader range of parameters because, even though the shapes of different data sets over the same energy range are similar, the cross-normalizations vary by 20-30\% for some data sets. However, it became clear that adopting choice (B) leads to formally allowed fits that fail to fit the data from a single instrument (e.g., the fit shown in Figure~\ref{fig:derosa-cxb} does not fit the Swift-BAT data). Even if the normalization of these data were wrong, the CXB fit within this data set can clearly be rejected. There are many examples like this, where the best-fit model cannot reproduce the shape of the CXB but fall within the uncertainty of cross-normalizations when we adopt choice (B). Therefore, we choose to focus only on the results for choice (A) errors. We note that, among these more recent data, the cross-normalizations are far less discrepant, supporting the idea that normalization errors do not dominate the analysis. The reader can investigate other choices using the interactive tool provided.

We apply a 3$\sigma$ threshold in Table~\ref{tab:sample_results}. Note that the spectral parameters measured in hard X-ray selected samples generally have higher P($\chi^2$), suggesting those AGN are more representative of the entire AGN population constituting the CXB. There are exceptions, however: for example, the hard X-ray selected local AGN sample of  \citet{dadina2008} has very high cutoff energies for both absorbed and unabsorbed objects, and produces poor fits to the CXB. Still, the lowest probabilities were found for soft X-ray-selected (E $<$ 10 keV) or Type~1-only samples, which are clearly less representative of the full AGN population. This is partly because some parameters such as cutoff energy and reflection scaling factor cannot be constrained using low energy windows. 

The reflection scaling factor does not seem to have any effect on the Compton-thick spectra, as shown in Figure~\ref{fig:gamma_variation_spectra}. This is expected, as Compton scattering out of line-of-sight (\textsc{cabs}) and photoelectric absorption (\textsc{tbabs}) significantly attenuate reflection from the accretion disk. As shown in Table~\ref{tab:spectral_parameters_in_papers}, some previous works suggest that reflection scaling factor decreases with absorption \citep{claudio2017bat,zap2018} even when attenuation is accounted for. \citet{beckmann2009} found that using the same value of R, but different inclination angles for different levels of absorption also fits observed spectra well. Therefore, it is possible that the observed decrease in R actually indicates that we are observing more obscured sources at higher inclination angles, and lends some support to the unified model of AGN. However, we did not force the unified model on the spectral sets that we assumed, and therefore, the inclination angle was kept fixed at the default value of 60$^{\circ}$ for all absorption bins.

For computational efficiency, we imposed more liberal constraints using only the necessary condition of reproducing the CXB. Fitting the number counts as well would further limit the spectral parameter space and can potentially be done in a future work. Here, we can qualitatively use number counts to rule out some regions of the parameter space which we included in our analysis because they are observed in some survey samples, but are unlikely to be representative of the full AGN population. Figure~\ref{fig:neural_network_result} shows that some good fits to the CXB exists for very high values of $\langle R \rangle$ (i.e.,  $\langle R \rangle= 1.5$). As shown in Figure~\ref{fig:gamma_variation_spectra}, a higher $\langle R \rangle$ contributes the most in the Compton hump region. Therefore, to produce a good fit to CXB, we need to lower the space densities of Compton-thick objects. Compton-thick number counts obtained by \citet{lansbury2017Ctk}, \citet{lanzuisi2018cosmosctk} predicts very high space densities of Compton-thick objects, so it is unlikely that $\langle R \rangle$ values much higher than 1.3 can fit the CXB and simultaneously fit such high space densities of Compton-thick objects.

Note that additional complexity may arise if some parameters are redshift and/or luminosity dependent. Currently, possible dependencies of the spectral parameters on luminosity and redshift are not well constrained. Therefore, we assumed constant values in order to be conservative. We leave the exploration of these aspects to future work.

\acknowledgements
The authors wish to thank the referee for suggesting substantial improvements to the representation of the quantitative analysis presented in this work. This material is based upon work supported by the National Science Foundation under Grant No.  AST-1715512, and Yale University. ET acknowledges support from FONDECYT Regular 1160999, FONDECYT Regular 1190818, CONICYT PIA ACT172033 and Basal-CATA PFB-06/2007 and AFB170002 grants. CR acknowledges the CONICYT PIA Convocatoria Nacional subvencion a instalacion en la academia convocatoria a\~{n}o 2017 PAI77170080. TA wishes to thank her family members: M. A. Quayum, Shamim
Ara Begum, Mahbubul Hoque Bhuiyan, Mehrab Bakhtiar, Arnita Tasnim Ankur and Raysa Tasnim Orin for their
support.  TA and CMU wishes to thanks Professor John Hartigan for help with the statistical analysis presented in this work. TA also wishes to thank Dr. Lia Sartori for advice on X-ray spectra.

\textit{Software:} {\tt numpy} \citep{numpy2011}, {\tt Astropy} \citep{astropy2018}, {\tt Emcee} \citep{emcee}
{\tt Matplotlib} \citep{matplotlib}, {\tt Topcat} \citep{taylor2005}, \textsc{xpec} and \textsc{pyxpsec} \citep{xspec}, {\tt ChainConsumer} \citep{chainconsumer} and {\tt Vegas} \citep{vegas}. Parts of the neural network code was adapted from \citet{nielsen2015} e-book.

	
\end{document}